\documentclass{aa}  

\usepackage{graphicx}
\usepackage{natbib}

\usepackage{booktabs}  
\usepackage{txfonts}

\graphicspath{{./Pictures/}} 

\begin{document}

   \title{HADES RV Programme with HARPS-N at TNG}

   \subtitle{X. The non-saturated regime of the stellar activity-rotation relationship for M-Dwarfs \thanks{Based on observations collected at the Italian Telescopio Nazionale Galileo (TNG), operated on the island of La Palma by the Fundación
Galileo Galilei of the INAF (Istituto Nazionale di Astrofisica) at the
Spanish Observatorio del Roque de los Muchachos of the Instituto
de Astrofísica de Canarias, in the framework of the The HArps-n red Dwarf Exoplanet Survey (HADES) observing program.}}

   \author{E. Gonz\'alez-\'Alvarez 
          \inst{1}
          \and G. Micela
          \inst{1}
          \and J. Maldonado
          \inst{1}
		  \and L. Affer
          \inst{1}
          \and A. Maggio
          \inst{1}   
          \and A. F. Lanza
          \inst{2}
          \and E. Covino
          \inst{3}       
          \and S. Benatti
          \inst{4}
          \and A. Bignamini
          \inst{5}      
          \and R. Cosentino
          \inst{6}
          \and M. Damasso
          \inst{7}      
          \and S. Desidera
          \inst{4}
          \and J. I. Gonz\'alez Hern\'andez
          \inst{8,9}
          \and A. Mart\'inez-Fiorenzano
          \inst{6}   
          \and I. Pagano
          \inst{2}
          \and M. Perger
          \inst{10,12}  
          \and G. Piotto
          \inst{4,11}   
          \and M. Pinamonti
          \inst{7}
          \and M. Rainer
          \inst{13}
          \and R. Rebolo
          \inst{8,9}
          \and I. Ribas
          \inst{10,12}
          \and G. Scandariato
          \inst{2}
          \and A. Sozzetti
          \inst{7}
          \and A. Su{\'a}rez Mascare{\~n}o
          \inst{14}
          \and B. Toledo-Padr\'on
          \inst{8,9}
          }

             \institute{INAF-Osservatorio Astronomico di Palermo,
              Piazza Parlamento 1, 90134 Palermo, Italy
			\and INAF-Osservatorio Astrofisico di Catania, via S. Sofia 78, 95123 Catania, Italia
			\and INAF-Osservatorio Astronomico di Capodimonte, Salita Moiariello 16, 80131 Napoli, Italia
			  \and INAF-Osservatorio Astronomico di Padua, Vicolo dell’Osservatorio 5, 35122 Padova, Italia
			  \and INAF-Osservatorio Astronomico di Trieste, via Tiepolo 11, 34143 Trieste, Italia
			   \and Fundaci\'on Galileo Galilei-INAF, Rambla Jos\'e Ana Fernandez P\'erez 7, 38712 Bre\~na Baja, TF, Espa\~na
               \and INAF-Osservatorio Astrofisico di Torino, via Osservatorio 20, 10025 Pino Torinese, Italia            
               \and Instituto de Astrof\'isica de Canarias, 38205 La Laguna, Tenerife, Spain
			\and Universidad de La Laguna, Dpto. Astrof\'isica, 38206 La Laguna, Tenerife, Spain               
               \and Institut de Ci\`encies de l'Espai (IEEC-CSIC), Campus UAB, Carrer de Can Magrans s/n, 08193, Bellaterra, Spain   
               \and Dipartimento di Fisica e Astronomia G. Galilei, Universit\`a di Padova, Vicolo dell’Osservatorio 2, 35122 Padova, Italia
               \and Institut d'Estudis Espacials de Catalunya (IEEC), 08034 Barcelona, Spain               
               \and INAF-Osservatorio Astrofisico di Arcetri, Largo Enrico Fermi 5, 50125 Firenze, Italia
                \and Observatoire Astronomique de l'Universit\`e de Gen\`eve, 1290 Versoix, Switzerland
				}

   \offprints{E. Gonz\'alez - \'Alvarez \\ \email{egonzalez@astropa.inaf.it}}
   \date{Received 5 October 2018 / Accepted 10 February 2019}

 
  \abstract
   {}
   {We aim to extend the relationship between X-ray luminosity ($L_{x}$) and rotation period ($P_{\rm rot}$) found for main-sequence FGK stars and test whether  it also holds for early-M dwarfs, especially in the non-saturated regime ($L_{x} \propto P_{\rm rot}^{-2}$) which corresponds to slow rotators.}
   {We use the luminosity coronal activity indicator ($L_{x}$) of a sample of 78 early-M dwarfs with masses in the range from 0.3 to 0.75 $M_{\sun}$ from the HArps-N red Dwarf Exoplanet Survey (HADES) radial velocity (RV) programme collected from $ROSAT$ and $XMM$-$Newton$. The determination of the rotation periods ($P_{\rm rot}$) was done by analysing time-series of high-resolution spectroscopy of the Ca~{\sc ii} H \& K and H$\rm \alpha$ activity indicators. Our sample principally covers the slow rotation regime with rotation periods from 15 to 60 days.}
   {Our work extends to the low mass regime the observed trend for more massive stars showing a continuous shift of the $L_{x}/L_{\rm bol}$ vs. $P_{\rm rot}$ power-law towards longer rotation period values and includes the determination, in a more accurate way, of the value of the rotation period at which the saturation occurs ($P_{\rm sat}$) for M dwarf stars.}
   {We conclude that the relations between coronal activity and stellar rotation for FGK stars also hold for early-M dwarfs in the non-saturated regime, indicating that the rotation period is sufficient to determine the ratio $L_{x}/L_{\rm bol}$. }


\maketitle
%

\section{Introduction}
\label{Introduction}

Currently, the search for small, rocky planets with the potential capability of hosting life has been focused around M dwarf stars. The Kepler mission has recently shown that terrestrial planets are more frequent around M dwarfs compared to solar-like FGK stars \citep{2012ApJS..201...15H,2013ApJ...767...95D}, so in the last years M dwarfs became more interesting targets for the search of planets.

Understanding the stellar activity is crucial to correctly interpret the physics of stellar atmospheres and the radial velocity data from ongoing exoplanet search programmes. This is particularly critical for M dwarfs as they show high stellar activity levels \citep[both chromospheric and coronal emission, e.g.][]{2007AcA....57..149K} due to their deep convective layers and they are on average more active than solar-like stars \citep[e.g. ][]{1997A&A...327.1114L,2005ApJ...621..398O}. Therefore we need to analyse very carefully the activity properties of these targets. Therefore we make a brief summary regarding the principal characteristics of M dwarfs which directly affect us in our study.

These stars are relatively cool (approx. between 2200 and 4000 K), have masses smaller than $ \sim 0.6 M_ \odot $ and offer several advantages. First of all, M dwarf stars are the most abundant component of the solar neighbourhood, $\sim 75 \%$ of the stars within 10 pc \citep[e.g.][]{2006AJ....132.2360H, 2002AJ....124.2721R}. Moreover, the contrast planet-star is more favorable: the motion induced by an Earth mass planet in the habitable zone around a M dwarf star is of the order of 1 $\rm m~s^{-1}$ (within today capabilities) while the same planet would induce a motion $\sim$ 10 $\rm cm~s^{-1}$ around solar-like star \citep[e.g.][]{2013ApJ...767...95D, 2013EPJWC..4703006S, 2012ApJS..201...15H}. Finally, M dwarfs are more likely to host rocky planetary companions \citep{2010ApJ...713..410B}. From an observational point of view, the chances of finding an Earth-like planet in the habitable zone of a star increase as the stellar mass and orbital period decrease \citep{1993Icar..101..108K}. As a consequence the small separation and shorter periods make the amplitude of the variation of RV large, the time scale of variation shorter and, therefore, the temporal stability of the instrument is less constraining. Of course the habitability is not guaranteed simply by assessing the distance from the star and several other factors such as stellar activity, may shift the habitable zone of the star \citep{2013A&A...557A..67V}. Stellar activity describes the various observational consequences of magnetic fields, which appear on the stellar photosphere, in the chromosphere or in the corona. All these phenomena affect the circumstellar environment including planets and we need to develop an optimal strategy to discern true keplerian signals from activity induced radial velocity (RV) variations to identify small rocky planets orbiting in the habitable zone of M dwarfs.

It is well known that for FGK stars activity and rotation are linked by the stellar dynamo and both decrease as the star ages \citep[e.g.][]{1981ApJ...248..279P,1990ApJS...72..191S,1990ApJS...74..891R,2012A&A...537A..94S}. Despite the increasing interest in M dwarfs, the activity of these stars is far from being fully understood, even if some studies suggest that the connection between age, rotation, and activity may also hold in early M dwarfs \citep[e.g.][]{1998A&A...331..581D,2003A&A...397..147P, 2017A&A...598A..27M}.

Magnetic activity in late-type main-sequence stars is an observable manifestation of the stellar magnetic fields and causes X-ray coronal emission which is stronger for more rapidly rotating stars. The generation of surface magnetic fields in solar-like stars is considered the end result of a complex dynamo mechanism, whose efficiency depends on the interaction between differential rotation and convection inside the star \citep{2013IAUS..294.....K}. Therefore, stellar rotation must play a very important role and numerous studies have searched for relations between several chromospheric ($H_{\rm \alpha}$ and Ca~{\sc ii} H \& K) \citep[e.g.][]{1984ApJ...279..763N,2003ApJ...583..451M,2015MNRAS.452.2745S,2017A&A...600A..13A,2017ApJ...834...85N} and coronal (X-ray) \citep[e.g.][]{1981ApJ...248..279P,1987ApJ...315..687M, 1996A&A...305..785R, 2003A&A...397..147P, 2007AcA....57..149K,2011ApJ...743...48W,2014ApJ...794..144R,2018MNRAS.479.2351W} magnetic activity indicators and stellar rotation rate. In a feedback mechanism, magnetic fields are responsible for the spin-evolution of the star \citep{2015ApJ...799L..23M}, and therefore rotation and magnetic fields are intimately linked and play a fundamental role in stellar evolution. Early works have used spectroscopic measurements of stellar rotation ($v \sin i$) with intrinsic ambiguities related to the unknown inclination angle and radius of the stars \citep{1981ApJ...248..279P}. Stellar rotation rates are best derived from the periodic brightness variations induced by cool star-spots (photometrically measured) which can be directly associated with the rotation period which has been proven more useful than $v \sin i$ \citep{2003A&A...397..147P, 2011ApJ...743...48W}.

Theory predicts a qualitative change of the dynamo mechanism at the transition into the fully convective regime \citep[spectral type $\sim$ M4,][]{2011ASPC..448..505S}. This makes studies of the rotation dependence of magnetic activity across the M spectral range crucial for understanding fully convective dynamos. Rotation-activity studies have been presented with different diagnostics for activity, as H$\alpha$ and X-ray emission. X-ray emission was shown to be more sensitive to low activity levels in M dwarfs \citep{2013MNRAS.431.2063S}. On the other hand, studies with optical emission lines (H$\alpha$, Ca~{\sc ii} H \& K) as activity indicators have been mostly coupled with $v \sin i$ as a rotation measure because both parameters can be obtained from the same set of spectra \citep{2010AJ....139..504B,2012AJ....143...93R}. The combination of H$\alpha$ data with photometrically measured M star rotation periods has been studied by \cite{2015ApJ...812....3W,2017ApJ...834...85N}.

In previous works \citep[e.g.][]{2003A&A...397..147P, 2016csss.confE..62S} the dependence between magnetic activity and rotation for M dwarf stars remained poorly constrained, specially in the non-saturated regime (slow rotators). Therefore the turn-over point between the saturated and non-saturated regimes and the slope of the decaying part of the relation was not well constrained. \cite{2016csss.confE..62S} studied a sample of 134 bright, nearby M dwarfs (spectral type K7-M6), where only a total of 26 stars had X-ray measurements in the archival data bases available at that time. These 26 stars were divided into three spectral type groups (K7-M2, M3-M4 and M5-M6). \cite{2003A&A...397..147P} studied a sample of 259 solar-type dwarfs in the $B-V$ range 0.5-2.0, all of them with X-ray counterpart. The sample was divided as a function of the stellar mass (8 groups from 0.22-1.29 $M_{\odot}$ range) in order to investigate how the observed spread of X-ray emission levels depends on the stellar mass and to determine the best-fit relations between X-ray emission and rotation period for each mass bin. 

\cite{2011ApJ...743...48W} extended the available dataset and presented a sample of 824 solar and late-type stars with X-ray luminosities and rotation periods in the mass range 1.16 - 0.09 $M_{\odot}$. Their sample in the M dwarf mass range was well populated down to low values. Analogously occurs in \cite{2014ApJ...794..144R} that used the same sample analyzed in \cite{2011ApJ...743...48W}.

Although our sample does not cover the regime in which the transition to a fully convective interior occurs, we consider that our sample can still provide useful information to constrain the coronal emission-rotation relationship for M dwarfs, especially because we sample the non-saturated regime. 

This work aims to test whether the relations between X-ray luminosity and stellar rotation investigated in literature for main-sequence FGK stars and for pre-main-sequence M stars also hold for the early-M dwarfs using our sample of 78 M dwarfs from HADES radial velocity programme. The work is structured as follows. In Sect. \ref{sec:The sample} we present the sample of stars used in this work. Section \ref{sec:Rotation periods} describes the determination of the rotation periods ($P_{\rm rot}$). Section \ref{sec:Coronal activity} explains in detail how we studied and obtained the coronal activity indicator X-ray luminosity ($L_{x}$), the comparison with the nearby stellar population to know how representative of the solar neighbourhood our M dwarf sample is, following with the study of rotation period-stellar activity connection in M dwarfs using our M stars sample. Finally, the summary and conclusions are presented in Sect. \ref{sec:Summary and conclusions}.

\section{The stellar sample}
\label{sec:The sample}

\begin{figure*}[!t]
\centering
\begin{minipage}{0.49\linewidth}
\includegraphics[angle=0,scale=0.56]{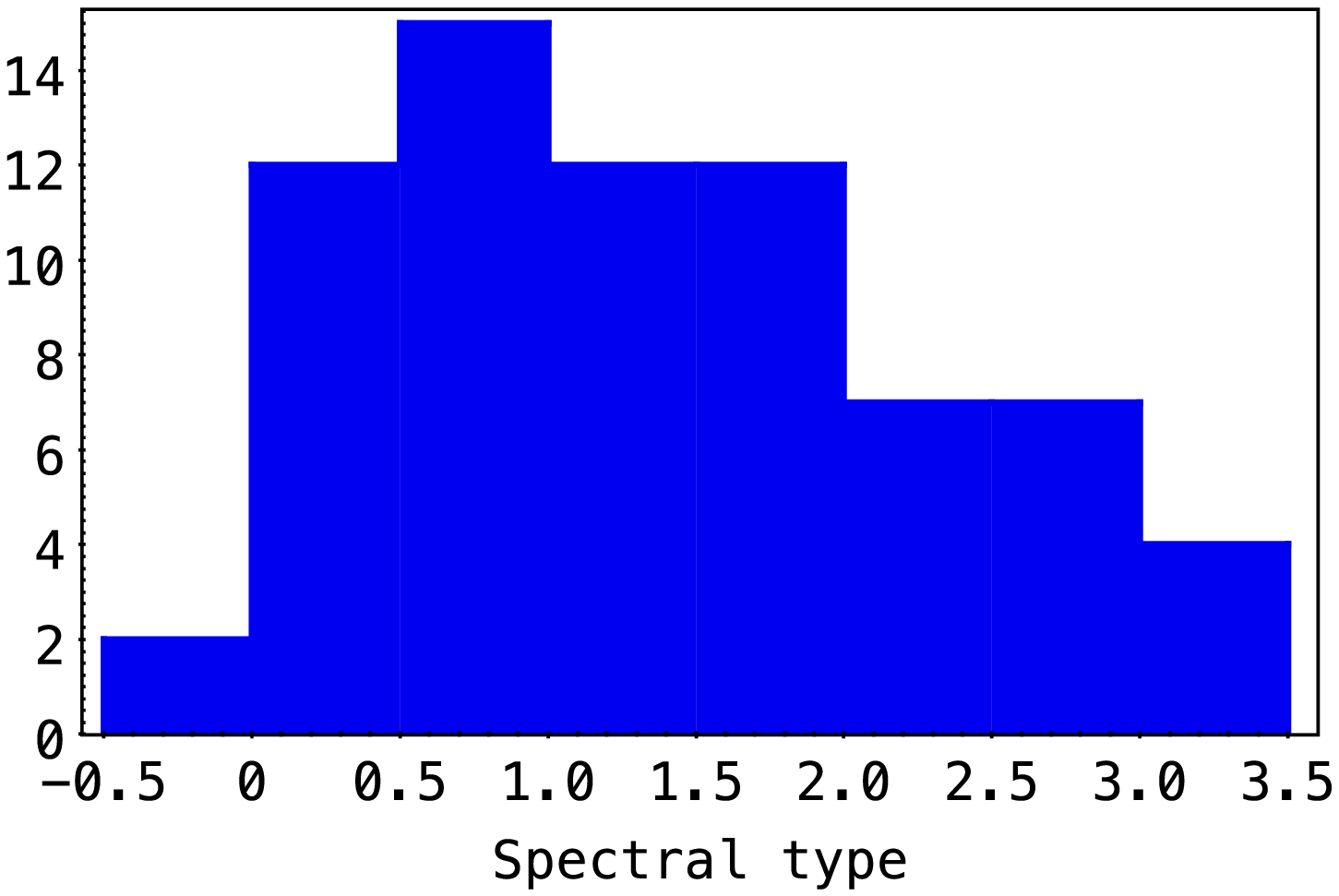}
\end{minipage}
\begin{minipage}{0.49\linewidth}
\includegraphics[angle=0,scale=0.56]{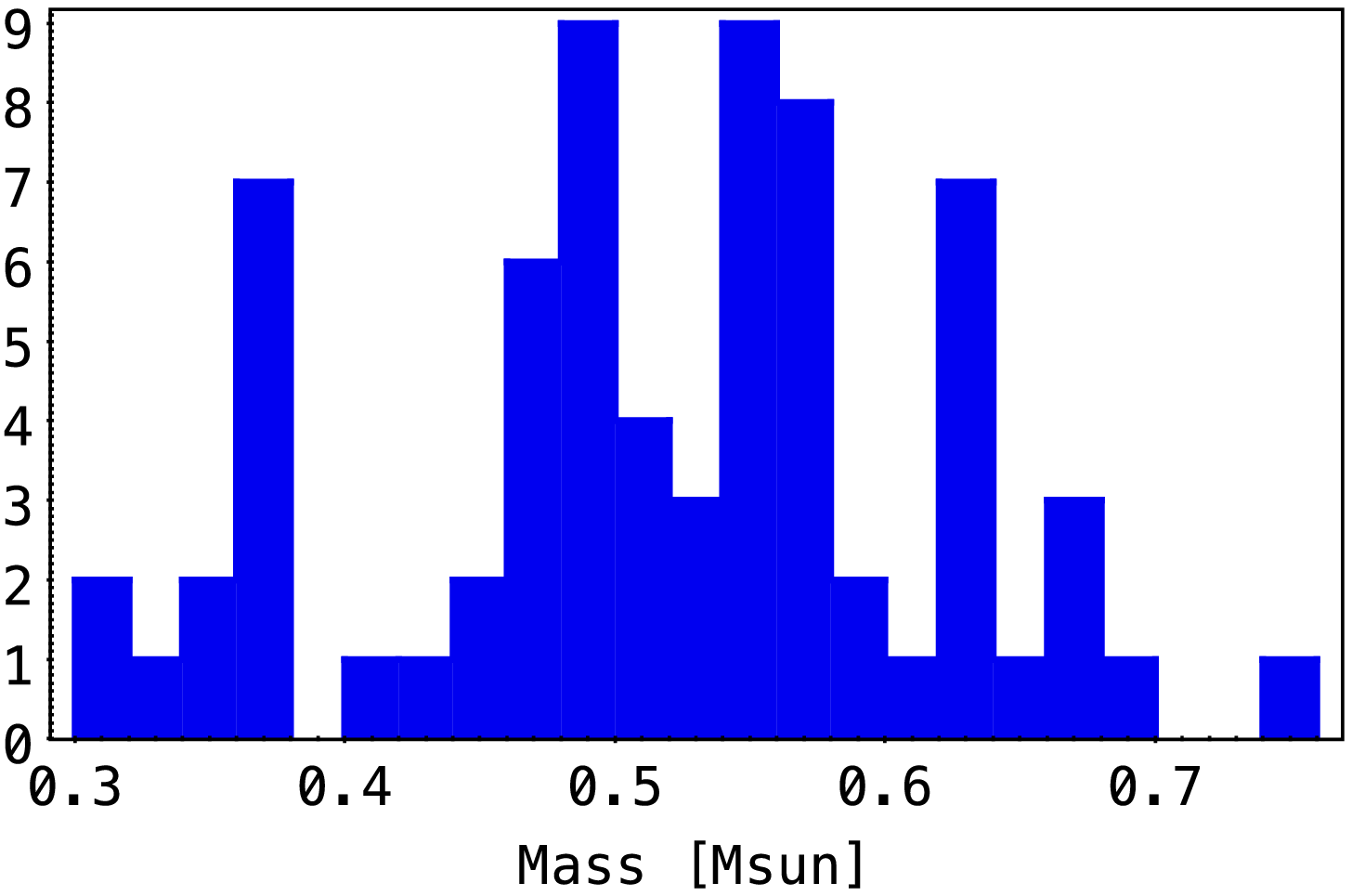}
\end{minipage}
\caption[Distribution of spectral types (on the left) and masses (on the right) for the HADES M dwarfs sample]{
Distribution of spectral types (on the left) and masses (on the right) for the HADES M dwarfs sample. Negative indices denote spectral types earlier than M, where the value -0.5 stands for K7.5.}
\label{hist_SpT_mass}
\end{figure*}

Our stellar sample was observed in the framework of the HArps-N red Dwarf Exoplanet Survey \citep[HADES,][]{2016A&A...593A.117A,2017A&A...598A..26P} a collaborative effort between the Global Architecture of Planetary Systems project \citep[GAPS,][]{2013A&A...554A..28C}, the Institut de Ci\`encies de l'Espai (ICE/CSIC) and the Instituto de Astrof\'isica de Canarias (IAC). The sample amounts to 78 late-K/early-M dwarfs and covers an effective temperature range from 3400 to 3900 K, corresponding to spectral types from K7.5 to M4V (Fig \ref{hist_SpT_mass}, left panel) and masses from about 0.3 to 0.75 $\rm M_{\odot}$ (Fig. \ref{hist_SpT_mass}, right panel). We group the stars in bins of 0.5 spectral subclasses, with K7.5 corresponding to -0.5, M0 to 0, and so on until M4, which is the last sub-type for which we have stars in our sample.

The stars were selected from the Palomar-Michigan State University (PMSU) catalogue \citep{1995AJ....110.1838R}, \cite{2011AJ....142..138L} and are targets observed by the APACHE transit survey \citep{2013EPJWC..4703006S} with a visible magnitude lower than 12. High-resolution \'echelle spectra of the stars were obtained at La Palma observatory (Canary Islands, Spain) during several observing runs between September 2012 and November 2017 using the HARPS-N instrument \citep{2012SPIE.8446E..1VC} at the Telescopio Nazionale Galileo (TNG). HARPS-N spectra cover the wavelength range 383-693 nm with a resolving power of R = 115,000 and all the spectra were automatically reduced using the Data Reduction Software \citep[DRS][]{2007A&A...468.1115L}. Basic stellar parameters (effective temperature, spectral type, surface gravity, iron abundance, mass, radius and luminosity) were computed and published by \cite{2017A&A...598A..27M} as part of the HADES collaboration using a methodology based on ratios of spectral features \cite{2015A&A...577A.132M}. The methodology consists of using ratios of pseudo-equivalent widths of spectral features as a temperature diagnostic. A list of calibrators was built for each of the basic stellar parameters considered  ($T_{\rm eff}$, spectral type, and metallicity) and the derived temperatures and metallicities were used together with photometric estimates of mass, radius, and surface gravity (more details in \cite{2015A&A...577A.132M}).

\section{Rotation periods}
\label{sec:Rotation periods}

\begin{figure}[!t]
\centering
\includegraphics[angle=0,scale=0.17]{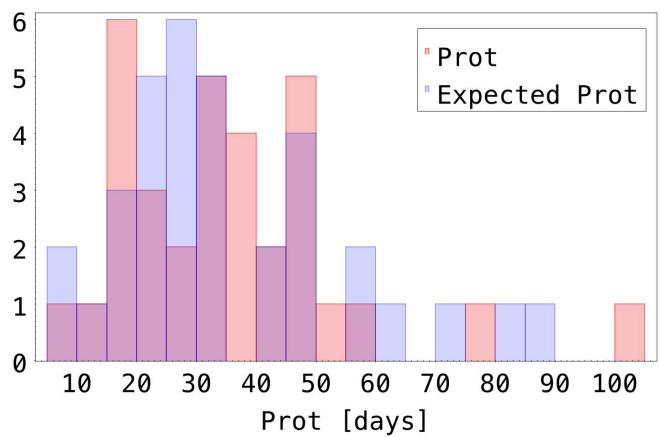}
\caption{Distribution of the rotation periods for the stars in our sample. The red area shows the measured rotation periods and the blue area corresponds to the derived rotation periods when a direct determination was not possible.}
\label{GAPS_hist_Prot}
\end{figure}

The determination of the rotation periods ($P_{\rm rot}$) was done by analysing time-series high-resolution spectroscopy of the Ca~{\sc ii} H \& K lines and H$\alpha$ activity indicators and was published in \cite{2018A&A...612A..89S}. The authors studied a fraction of our original M dwarf sample, composed of 72 stars (out of a total of 78) providing 33 rotation periods measured from the spectral indicator modulation as well as 34 periods derived from the Ca~{\sc ii} H \& K - rotational period relationship. The distribution of rotation periods is shown in Fig. \ref{GAPS_hist_Prot}. It can be seen that our $P_{\rm rot}$ values cover the range from 8 to 85 days. This interval exactly corresponds to a specific region in which planets in habitable zone around low-mass stars could be discovered. In other words, the period of planets in the habitable zone orbiting early-M dwarfs coincide with the stellar rotation period (as shown in Fig. 1 from \cite{2016ApJ...821L..19N}).

The vast majority of our studied sample covers the stellar rotation period range from 15 to 60 days and mass range to 0.3 to 0.75 $M_{\odot}$.

\section{Coronal activity}
\label{sec:Coronal activity}

As said in previous sections, traditionally the most frequently investigated diagnostics of chromospheric activity are the Ca~{\sc ii} H \& K and the Balmer lines measured from optical spectra. The $H_{\alpha}$ line has in general a complex structure, while Ca~{\sc ii} H \& K lines are difficult to be measured in red stars. In order to measure a stellar activity diagnostic independent of the optical spectra, we decided to investigate the coronal activity indicator X-ray luminosity ($L_x$). $L_x$ is emitted from the upper atmosphere and can be easily determined for nearby M stars.

With the purpose of studying the coronal activity, we collected the information for our sample of 78 M dwarfs from the main X-ray missions, $XMM$-$Newton$, $Chandra$ and $ROSAT$ \citep[e.g.][]{1992QJRAS..33..165T, 1999A&A...349..389V} following  this order of preference. Using The High Energy Astrophysics Science Archive Research Center (HEASARC) \footnote{\url{https://heasarc.gsfc.nasa.gov}} let us find what X-ray mission observed our selected targets or if they were observed by more than one mission. In that case, we followed the stablished preference. Even if $ROSAT$ is older and less sensitive than the most recent X-ray observatories, it is still the main source of X-ray measurements since it covered all the sky. In particular, the catalogues that provided us X-ray information were the $ROSAT$ Bright and Faint Source catalogues \citep[BSC and FSC:][]{1999A&A...349..389V}, the second $ROSAT$ All-Sky Survey Point Source Catalog \citep[2RXS:][]{2016A&A...588A.103B}, the third $ROSAT$ Catalogue of Nearby Stars \citep[CNS3:][]{1999A&AS..135..319H}, the $XMM$-$Newton$ XAssist source list \citep[XAssist:][]{2003ASPC..295..465P}, the $XMM$-$Newton$ Slew Survey Full Source Catalog, v2.0 \citep[XMMSLEW:][]{2008A&A...480..611S} and the Third $XMM$-$Newton$ Serendipitous Source Catalogue \citep[3XMM-DR7:][]{2016A&A...590A...1R}. 

For $ROSAT$ detected sources, we need to apply a count-to-flux Conversion Factor (CF) because the X-ray emission is expressed by count rates (CRT) and hardness ratio (HR) quantities. The HR is defined through $HR=\frac{H-S}{H+S}$, denoting by H and S the counts recorded in the soft (0.1-0.4 keV) and hard (0.5-2.0 keV) channels of the Position Sensitive Proportional Counters (PSPC) detector. In order to obtain the X-ray flux ($f_x$) from the PSPC detector, an appropriate conversion factor for a coronal spectral model must be computed \citep{1995ApJ...450..392S}:

\begin{equation}
CF=(8.31+5.30 \cdot HR)\times 10^{-12} ~ [\rm erg ~ cm^{-2}~ count^{-1}]
\end{equation}

The conversion factor depends on the given spectral model and on the instrument, the details on this conversion procedure are given by \cite{1995ApJS...99..701F}. Clearly, the conversion needs some assumptions for the intrinsic source spectrum, but fortunately the conversion factor does not depend sensitively on the adopted coronal temperature and also the absorption due to the interstellar material can be neglected due to the closeness of our stars. For the four cases with no hardness ratio (HR) in the database it is assumed a HR value of -0.4, corresponding to the middle activity level star \citep{1995ApJ...450..392S,1998A&AS..132..155H,2011A&A...532A...6S} and then used a fixed conversion factor value of $\rm 6.19 \times 10^{-12} ~erg~ cm^{-2}~ count^{-1}$. 

Combining the X-ray count rate with the conversion factor, the X-ray flux can be estimated as

\begin{equation}
 f_{x}=CF\cdot CRT ~ [\rm erg ~ cm^{-2} ~ s^{-1}]
 \label{ec_flujo}
 \end{equation}
where the error on the observed flux is determined by the signal-to-noise ratio ($S/N=\frac{eCRT}{CRT}$):
 
\begin{equation}
ef_{x}=\frac{eCRT}{CRT}f_{x}
\end{equation}

For $XMM$-$Newton$ detected sources (G243-30, GJ412A, GJ476, GJ49, GJ908, GJ9440, NLTT51676 and TYC2703-706-1) no conversion factor is needed because the X-ray emission is given directly on the observed flux with its correspondent error. Four of $XMM$-$Newton$ detected sources (GJ49, GJ9440, GJ908 and GJ412A) were collected with 3XMM-DR7 catalogue. Allowing us to calculate for these targets the X-ray flux in the same energy band used by $ROSAT$, using the Chandra Proposal Planning Toolkit PIMMS \footnote{\url{http://cxc.harvard.edu/toolkit/pimms.jsp}}. Assuming a plasma model (Astrophysical Plasma Emission Code, APEC), selecting the corresponding detector/filter used in the observation and setting the appropriate input (0.2-12 keV, the total energy band used in the 3XMM-DR7 processing) and output energy band (0.1-2.4 keV, the total energy band in the $ROSAT$ channels), the X-ray flux was derived. For the rest of sources detected by $XMM$-$Newton$ (G243-30, GJ476, NLTT51676, TYC2703-706-1) the different $XMM$-$Newton$ catalogues used (XMMSLEW and XAssist) did not allow us to use the previous approach with PIMMS due to a lack of information in the catalogue (e.g. detector/filter). Therefore, we provide directly the X-ray flux values reported by the corresponding catalogues (see XMM Science Survey Center memo SSC-LUX TN-0059 for a general description of the technique). We can estimate a systematic error introduced by the procedure described above of the order of 20-30\%. The XMM-Newton detected sources (G243-30, GJ476, NLTT51676 and TYC2703-706-1) where it was not possible to convert the X-ray flux in the same energy band of ROSAT, were not used in our analysis for two reasons: not having information on the rotation period or having a period derived from an activity-rotation relationship.

Once obtained the X-ray flux value available for our targets, the X-ray luminosity can be derived by

\begin{equation}
L_x=f_x4\pi d^2\,\,\,[erg\,s^{-1}]
\label{ec_luminosidad}
\end{equation} 

where $f_x$ corresponds to the X-ray observed flux and $d$ is the distance of the star. Table \ref{Tabla_emisionX} presents the 37 out of 78 stars with X-ray information collected for our M dwarfs sample. The total uncertainties on the derived luminosities are calculated by error propagation. In Table \ref{Tabla_emisionX} we also list the ratio $L_x/L_{\rm bol}$, where $L_{\rm bol}$ is the star bolometric luminosity, that is a measure of the coronal activity independent of the stellar size useful for a generalized analysis of the activity-rotation relation. The bolometric luminosity of the stars ($L_{\star}$ or $L_{\rm bol}$) was calculated following the Stefan-Boltzmann law:

\begin{equation}
L_{\star}=4 \pi R_{\star}^{2} \sigma_{B} T_{\rm eff}^4\,\,\,[erg\,s^{-1}]
\end{equation}
where $\sigma_{B}$ is the Stefan-Boltzmann constant ($5.6704 \cdot 10^{-5}$ $\rm erg~ cm^{-2}~s^{-1}~K^{-4}$). The $R_{\star}$ and $T_{\rm eff}$ are the radius and effective temperature of the star, respectively and were computed using the methodology based on ratios of spectral features described in \cite{2015A&A...577A.132M}. The corresponding stellar parameters for our M dwarfs sample were published in \cite{2017A&A...598A..27M}.

\begin{table}
\caption{\label{Tabla_emisionX} Stars with derived X-ray emission.}
\centering
\begin{tabular}{{p{0.28\linewidth}ccc}}
\noalign{\smallskip}
\hline\hline
\noalign{\smallskip}
 Name 		& 	$\log L_x$  		& $\log L_x/L_{\rm bol}$  \\
	  &	$[\rm erg ~ s^{-1}]$ &  	\\
\noalign{\smallskip}	
\hline
\noalign{\smallskip}

G243-30        &        $  28.87       \pm     0.13$   &  $ -3.16 \pm  0.43$ \\ 
GJ15A          &          $  27.29       \pm     0.05$   &  $ -4.64 \pm 0.12 $ \\   
GJ2            &        $  27.56       \pm     0.13$   &  $ -4.63 \pm 0.16 $ \\   
GJ2128         &        $  27.60       \pm     0.18$   &  $ -4.20 \pm 0.24 $ \\
GJ26           &        $  27.18       \pm     0.17$   &  $ -4.66 \pm 0.23 $ \\	
GJ272          &        $  27.38       \pm     0.21$   &  $ -4.71 \pm 0.23 $ \\
GJ3014         &        $  27.95       \pm     0.11$   &  $ -4.20 \pm 0.15 $ \\
GJ3117A        &        $  27.23       \pm     0.19$   &  $ -4.78 \pm 0.21 $ \\
GJ3822         &        $  27.76       \pm     0.19$   &  $ -4.59 \pm 0.22 $ \\
GJ3942         &        $  27.68       \pm     0.17$   &  $ -4.66 \pm 0.20 $ \\
GJ408          &        $  26.93       \pm     0.21$   &  $ -4.86 \pm 0.26 $ \\
GJ412A         &        $  26.32       \pm     0.03$   &  $ -5.62 \pm 0.12 $ \\
GJ414B         &        $  27.52       \pm     0.23$   &  $ -4.65 \pm 0.24 $ \\
GJ450          &        $  27.65       \pm     0.12$   &  $ -4.44 \pm 0.16 $ \\
GJ47           &        $  27.27       \pm     0.22$   &  $ -4.59 \pm 0.27 $ \\
GJ476          &        $  27.05       \pm     0.15$   &  $ -4.82 \pm 0.22 $ \\
GJ49 		&        $  27.57       \pm     0.02$   &  $ -4.70 \pm 0.09 $ \\
GJ548A         &        $  28.04       \pm     0.11$   &  $ -4.43 \pm 0.15 $ \\
GJ552          &        $  27.58       \pm     0.19$   &  $ -4.50 \pm 0.21 $ \\
GJ606          &        $  27.72       \pm     0.16$   &  $ -4.37 \pm 0.19 $ \\
GJ625          &        $  26.87       \pm     0.03$   &  $ -4.99 \pm 0.24 $ \\
GJ685          &        $  27.45       \pm     0.07$   &  $ -4.87 \pm 0.12 $ \\
GJ694.2        &        $  27.48       \pm     0.19$   &  $ -4.70 \pm 0.21 $ \\
GJ70           &        $  27.41       \pm     0.03$   &  $ -4.46 \pm 0.14 $ \\
GJ720A         &        $  27.39       \pm     0.15$   &  $ -5.11 \pm 0.18 $ \\
GJ740          &        $  27.51       \pm     0.14$   &  $ -4.85 \pm 0.17 $ \\
GJ793          &        $  27.77       \pm     0.03$   &  $ -3.99 \pm 0.18 $ \\
GJ835	       &        $  27.97       \pm     0.12$   &  $ -4.27 \pm 0.43 $ \\
GJ895          &        $  27.20       \pm     0.21$   &  $ -5.07 \pm 0.23 $ \\
GJ908          &        $  26.60       \pm     0.07$   &  $ -5.28 \pm 0.14 $ \\
GJ9440         &        $  27.00       \pm     0.05$   &  $ -5.24 \pm 0.10 $ \\
GJ9793         &        $  28.80       \pm     0.18$   &  $ -3.82 \pm 0.24 $ \\
NLTT51676      &        $  29.29       \pm     0.22$   &  $ -3.09 \pm 0.43 $ \\
NLTT53166      &        $  28.58       \pm     0.08$   &  $ -3.80 \pm 0.12 $ \\
TYC2703-706-1  &        $  29.97       \pm     0.12$   &  $ -2.48 \pm 0.15 $ \\
TYC2710-691-1  &        $  28.89       \pm     0.12$   &  $ -3.61 \pm 0.15 $ \\
TYC3720-426-1  &          $ 29.18        \pm     0.08$   &  $ -3.29 \pm 0.17 $ \\

\noalign{\smallskip}
\hline
\end{tabular}
\end{table}

\subsection{Comparison with the nearby stellar population}
\label{subsec:Lx population}

\begin{figure}[!t]
\centering
\includegraphics[angle=0,scale=0.60]{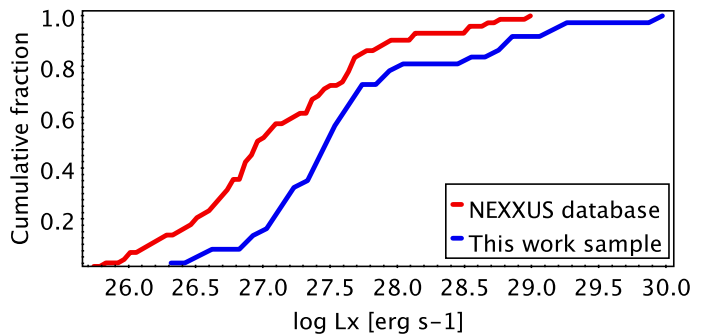}
\caption[Cumulative distribution function of $\log L_x$]{  
Cumulative distribution function of $\log L_x$. The red and blue lines correspond with the NEXXUS M dwarfs and this work sample, respectively.}
\label{CF_NEXXUS_GAPS_sample}
\end{figure}

In order to test how representative is our M dwarf sample of the solar neighbourhood, we compare its X-ray properties with a large sample of nearby M stars. The comparison sample is taken from the NEXXUS database \citep{2004A&A...417..651S} where only M dwarfs between M0 to M4.5 spectral types were selected (the same spectral type range of our sample). NEXXUS is a catalogue of all known stars within a distance of 25 pc to the Sun that are identified as X-ray and/or XUV-emitting stars from $ROSAT$ data, based on positional coincidence. Figure \ref{CF_NEXXUS_GAPS_sample} shows the cumulative distribution functions of $\log L_x$ for the NEXXUS M dwarfs and for our sample. The comparison shows a clear tendency of our sample to show higher levels of activity in X-ray emission with a $\log L_x$ median value about twice the median value of the nearby population. A Kolmogorov-Smirnov (K-S) test shows the significant difference between the two samples with a K-S statistic value $D$=0.41 and $p$-value=$0.0002$. We collected X-ray information close to the 50\% of the total sample. Therefore, as a roughly approximation, we are studying the regime of moderately active M dwarfs and our research for exoplanets may be made more difficult by the presence of stellar activity.

\subsection{Rotation period - activity relationship}

In order to study the M star rotation-activity connection, we use the X-ray data collected in Sect. \ref{sec:Coronal activity} and the rotation periods determined in the framework of the HADES collaboration and explained in Sect. \ref{sec:Rotation periods}. Merging both quantities, our sample was reduced to 33 stars with both measurements. The final sample is given in Table \ref{tab:x_prot} (a total of 19 measured $P_{\rm rot}$ from the spectral indicators) and Table \ref{tab:x_p-exp} (a total of 14 derived $P_{\rm rot}$ from the Ca~{\sc ii} H \& K - $P_{\rm rot}$ relationship).

\begin{small}
\begin{table*}
\caption{\label{tab:x_prot} Stars with X-ray activity level and \textit{measured} rotation periods. Rotation values are from \cite{2018A&A...612A..89S} while $L_x$ and $L_{\rm bol}$ values are computed in this work.}
\centering
\begin{tabular}{lccccc}
\noalign{\smallskip}
\hline\hline
\noalign{\smallskip}
Name		& Spectral type*	& $T_{\rm eff}$	&$\log L_x$	& $\log L_x/L_{\rm bol}$		&	 $P_{rot}$\\

		&		&	(K) & ($\rm erg~s^{-1}$)	&	& (days)\\
\noalign{\smallskip}
\hline
\noalign{\smallskip}

GJ3942       & 0.0 & $ 3867 \pm 69    $ & $27.80 \pm 0.18$ & $-4.66 \pm 0.20$ & $16.3 \pm 0.1$ \\
GJ548A       & 0.0 & $ 3903 \pm 70    $ & $28.04 \pm 0.11$ & $-4.43 \pm 0.15$ & $36.6 \pm 0.1$ \\
TYC2703-706-1& 0.5 & $ 3822 \pm 70    $ & $29.97 \pm 0.12$ & $-2.48 \pm 0.15$ & $7.8  \pm 0.2$ \\  
GJ685        & 0.5 & $ 3816 \pm 69    $ & $27.45 \pm 0.07$ & $-4.87 \pm 0.11$ & $16.3 \pm 4.2$ \\ 
GJ720A       & 0.5 & $ 3837 \pm 69    $ & $27.26 \pm 0.15$ & $-5.11 \pm 0.18$ & $34.5 \pm 4.7$ \\  
GJ412A       & 0.5 & $ 3631 \pm 68    $ & $26.32 \pm 0.03$ & $-5.62 \pm 0.12$ & $100.9\pm 0.3$ \\  
GJ694.2      & 0.5 & $ 3847 \pm 69    $ & $27.64 \pm 0.18$ & $-4.70 \pm 0.21$ & $17.3 \pm 0.1$ \\  
GJ740        & 0.5 & $ 3845 \pm 69    $ & $27.52 \pm 0.15$ & $-4.85 \pm 0.17$ & $36.4 \pm 1.7$ \\  
GJ3822       & 0.0 & $ 3821 \pm 70    $ & $27.76 \pm 0.20$ & $-4.59 \pm 0.22$ & $18.3 \pm 0.1$ \\  
GJ2          & 1.0 & $ 3713 \pm 68    $ & $27.57 \pm 0.13$ & $-4.63 \pm 0.16$ & $21.2 \pm 0.5$ \\
GJ15A        & 1.0 & $ 3607 \pm 68    $ & $27.29 \pm 0.05$ & $-4.64 \pm 0.12$ & $45.0 \pm 4.4$ \\
GJ49	      & 1.5 & $ 3712 \pm 68    $ & $27.57 \pm 0.02$ & $-4.70 \pm 0.09$ & $18.4  \pm 0.7$ \\
GJ908        & 1.5 & $ 3553 \pm 68    $ & $26.60 \pm 0.07$ & $-5.28 \pm 0.14$ & $49.9 \pm 3.5$ \\ 
GJ9440       & 1.5 & $ 3710 \pm 68    $ & $27.00 \pm 0.05$ & $-5.24 \pm 0.10$ & $48.0 \pm 4.8$ \\ 
GJ606        & 1.5 & $ 3665 \pm 68    $ & $27.72 \pm 0.16$ & $-4.37 \pm 0.19$ & $20.0 \pm 2.0$ \\  
GJ47         & 2.0 & $ 3525 \pm 68    $ & $27.27 \pm 0.22$ & $-4.59 \pm 0.27$ & $34.7 \pm 0.1$ \\ 
GJ625        & 2.0 & $ 3499 \pm 68    $ & $26.70 \pm 0.16$ & $-4.99 \pm 0.24$ & $77.8 \pm 5.5$ \\  
GJ552        & 2.0 & $ 3589 \pm 68    $ & $27.58 \pm 0.19$ & $-4.50 \pm 0.21$ & $43.5 \pm 0.1$ \\  
GJ476        & 3.0 & $ 3498 \pm 69    $ & $27.05 \pm 0.15$ & $-4.82 \pm 0.22$ & $55.0 \pm 5.5$ \\ 

\noalign{\smallskip}
\hline
\end{tabular}
\tablefoot{$^{*}$ Spectral type = 0 - 3.0 correspond to M0 - M3.0 spectral types.}
\end{table*}
\end{small}


\begin{small}
\begin{table*}
\caption{\label{tab:x_p-exp} Stars with X-ray activity level and \textit{derived} rotation periods. Rotation values are from \cite{2018A&A...612A..89S} while $L_x$ and $L_{\rm bol}$ values are computed in this work.}
\centering
\begin{tabular}{lccccc}
\noalign{\smallskip}
\hline\hline
\noalign{\smallskip}
Name		& Spectral type*	& $T_{\rm eff}$  &	$\log L_x$	& $\log L_x/L_{\rm bol}$		&	Derived $P_{rot}$\\

		&		&	(K) & ($\rm erg~s^{-1}$)	&	& (days)\\
\noalign{\smallskip}
\hline
\noalign{\smallskip}

TYC2710-691-1& -0.5 & $ 3867 \pm 71 $ &  $28.89 \pm 0.12$ & $-3.61 \pm 0.15$ &  $34.0  \pm        6.0 $   \\
GJ9793       &  0.0 & $ 3881 \pm 70 $ &  $28.80 \pm 0.18$ & $-3.82 \pm 0.24$ &  $ 15.0 \pm        3.0 $   \\
NLTT53166    &  0.0 & $ 3832 \pm 70 $ &  $28.58 \pm 0.08$ & $-3.80 \pm 0.12$ &  $ 55.0 \pm        9.0 $   \\
TYC3720-426-1&  0.0 & $ 3822 \pm 70 $ &  $29.18 \pm 0.08$ & $-3.29 \pm 0.17$ &  $8.0   \pm        1.0 $   \\
GJ272        & 1.0  & $ 3747 \pm 68 $ &  $27.50 \pm 0.21$ & $-4.71 \pm 0.23$ &  $41.0  \pm        7.0 $   \\
GJ895        & 1.5  & $ 3748 \pm 68 $ &  $27.20 \pm 0.21$ & $-5.07 \pm 0.23$ &  $24.0  \pm        5.0 $   \\
GJ3014       & 1.5  & $ 3695 \pm 69 $ &  $27.95 \pm 0.11$ & $-4.20 \pm 0.15$ &  $24.0  \pm        5.0 $   \\
GJ414B       & 2.0  & $ 3661 \pm 68 $ &  $27.52 \pm 0.23$ & $-4.65 \pm 0.24$ &  $62.0  \pm        10.0$   \\
GJ3117A      & 2.5  & $ 3549 \pm 68 $ &  $27.23 \pm 0.19$ & $-4.78 \pm 0.21$ &  $22.0  \pm        4.0 $   \\
GJ70         & 2.5  & $ 3511 \pm 68 $ &  $27.41 \pm 0.03$ & $-4.46 \pm 0.14$ &  $46.0  \pm        8.0 $   \\
GJ26         & 2.5  & $ 3484 \pm 68 $ &  $27.18 \pm 0.17$ & $-4.66 \pm 0.23$ &  $27.0  \pm        5.0 $   \\
GJ408        & 2.5  & $ 3472 \pm 68 $ &  $26.93 \pm 0.21$ & $-4.86 \pm 0.26$ &  $58.0  \pm        10.0$   \\
GJ2128       & 2.5  & $ 3518 \pm 68 $ &  $27.60 \pm 0.18$ & $-4.21 \pm 0.24$ &  $85.0  \pm        15.0$   \\
GJ793        & 3.0  & $ 3461 \pm 68 $ &  $27.77 \pm 0.03$ & $-3.99 \pm 0.18$ &  $34.0  \pm        6.0 $   \\ 

\noalign{\smallskip}
\hline
\end{tabular}
\tablefoot{$^{*}$ Spectral type = -0.5 - 3.0 correspond to K7.5 - M3.0 spectral types.}
\end{table*}
\end{small}

\begin{figure*}[!htp]
\centering
\begin{minipage}{0.48\linewidth}
\includegraphics[angle=0,scale=0.5]{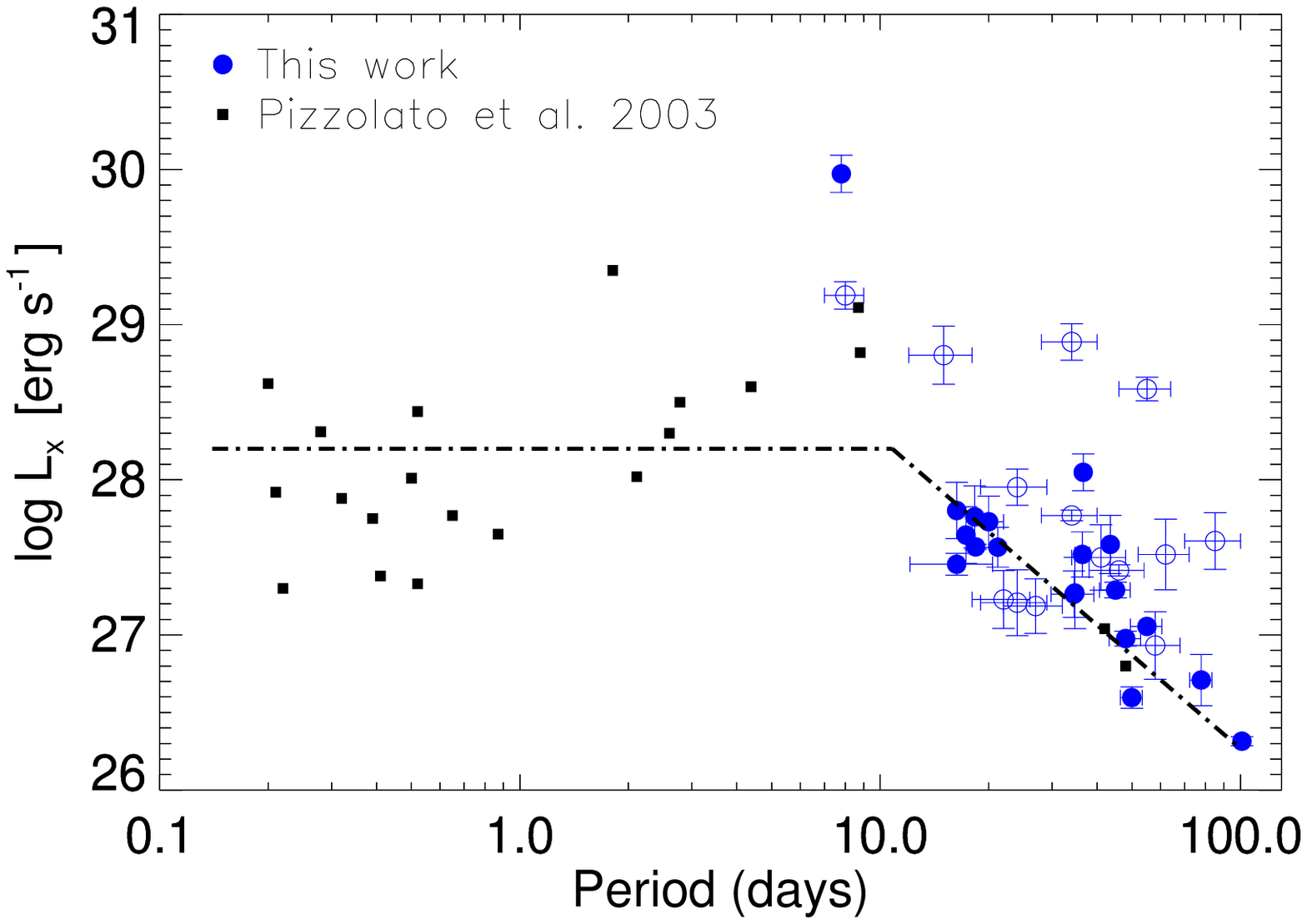}
\end{minipage}
\begin{minipage}{0.48\linewidth}
\includegraphics[angle=0,scale=0.5]{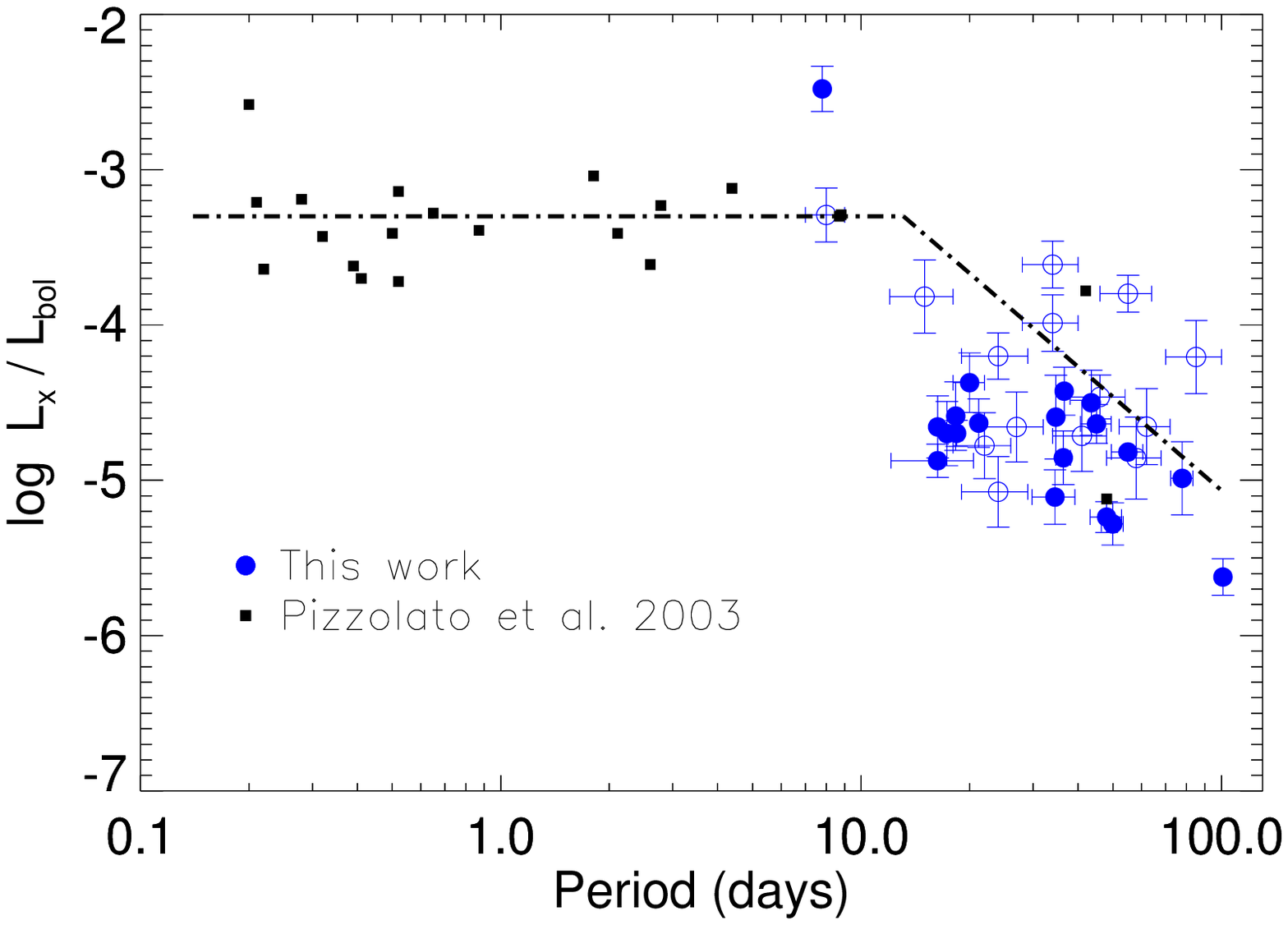}
\end{minipage}
\caption[Activity-rotation relationship]{
$L_x$ (Left panel) and $L_x/L_{\rm bol}$ (right panel) versus rotation period. The black squares correspond to the stellar sample from \cite{2003A&A...397..147P} in the mass range 0.22 < $M/M_{\odot}$ < 0.60. The blue large dots correspond to the M dwarfs with periods measured from time series and used in the analysis. The blue open circles show the derived period from an activity-rotation relation (excluded from the analysis). The black dot-dashed line represents the broken power law obtained by the fitting procedure from \cite{2003A&A...397..147P} with $P_{\rm rot}$ values estimated from $v \sin i$ in saturated regime.}
\label{Lx_and_Lbol_prot_all}
\end{figure*}

The mass range of our sample of M dwarfs (0.3 to 0.75 $M_{\odot}$) is comparable with the M dwarfs mass range of previous studies. In Fig. \ref{Lx_and_Lbol_prot_all} we present an update of the activity-rotation relation of M dwarf stars using the X-ray data extracted from archives and the rotation periods by \cite{2018A&A...612A..89S}, including stars with  rotation periods derived from the activity-rotation relationship that are plotted as open circles (see below). We also indicate the relation derived by \cite{2003A&A...397..147P} and their sample of 21 stars for their lowest mass bin, $M$ = 0.22-0.60 $\rm M_{\odot}$. It must be noted that the derived relation in the saturated regime by \cite{2003A&A...397..147P} is dominated by $P_{\rm rot}$ values estimated from $v \sin i$ measurements (therefore only upper limits values for $P_{\rm rot}$) and only two stars are present in the non-saturated regime. On the other hand, most of our data fill the slow rotation area of low-mass stars covering the regime poorly populated in previous studies \citep[e.g.][]{2003A&A...397..147P, 2016csss.confE..62S}.

\begin{figure*}[!htp]
\centering
\begin{minipage}{0.48\linewidth}
\includegraphics[angle=0,scale=0.5]{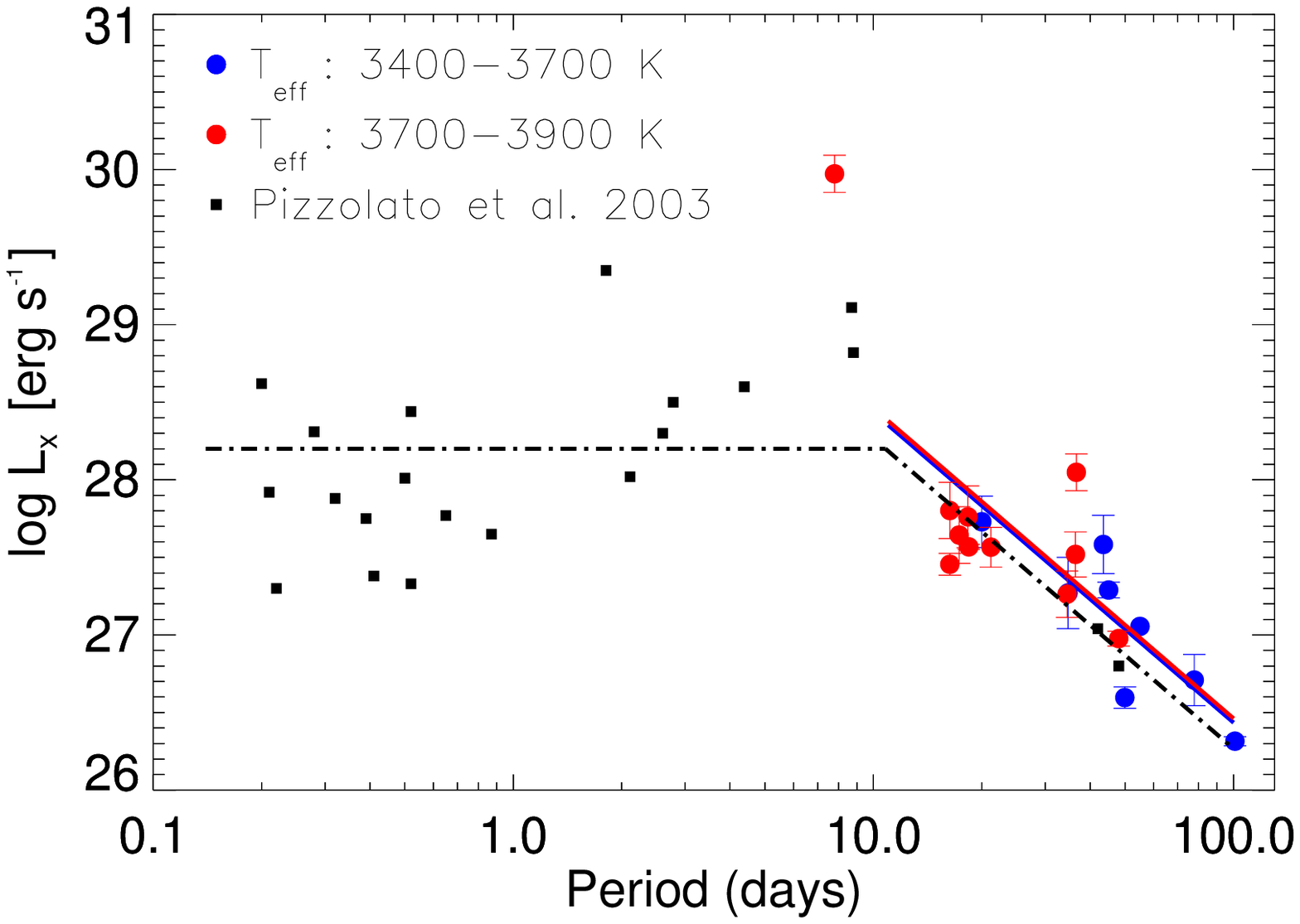}
\end{minipage}
\begin{minipage}{0.48\linewidth}
\includegraphics[angle=0,scale=0.5]{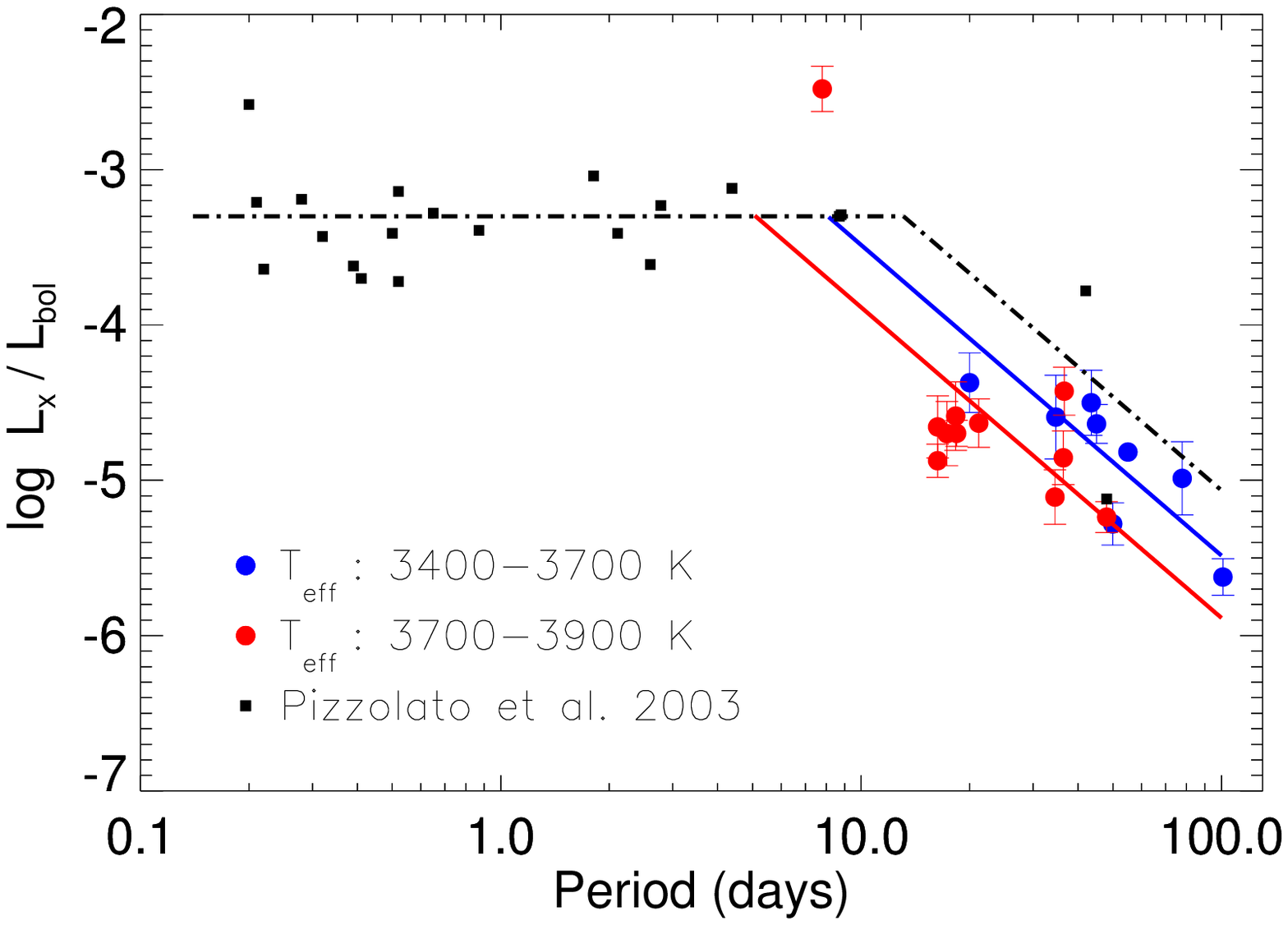}
\end{minipage}
\caption[Activity-rotation relationship as a function of the temperature.]{$L_x$ (left) and $L_x/L_{\rm bol}$ (right) versus rotation period for stars with direct rotation period determination. The black squares correspond to the stellar sample from \cite{2003A&A...397..147P} in the mass range 0.22 < $M/M_{\odot}$ < 0.60. Blue and red dots correspond to the M dwarfs from this work in the two different ranges of $T_{\rm eff}$ indicated in the legend. The black dot-dashed line represents the broken power law obtained by the fitting procedure from \cite{2003A&A...397..147P}. The blue and red solid lines represent our best fit for the 3400-3700 K and 3700-3900 K $T_{\rm eff}$ range, respectively.}
\label{Lx_and_Lbol_vs_prot}
\end{figure*}

\begin{figure*}[!htp]
\centering
\begin{minipage}{0.48\linewidth}
\includegraphics[angle=0,scale=0.5]{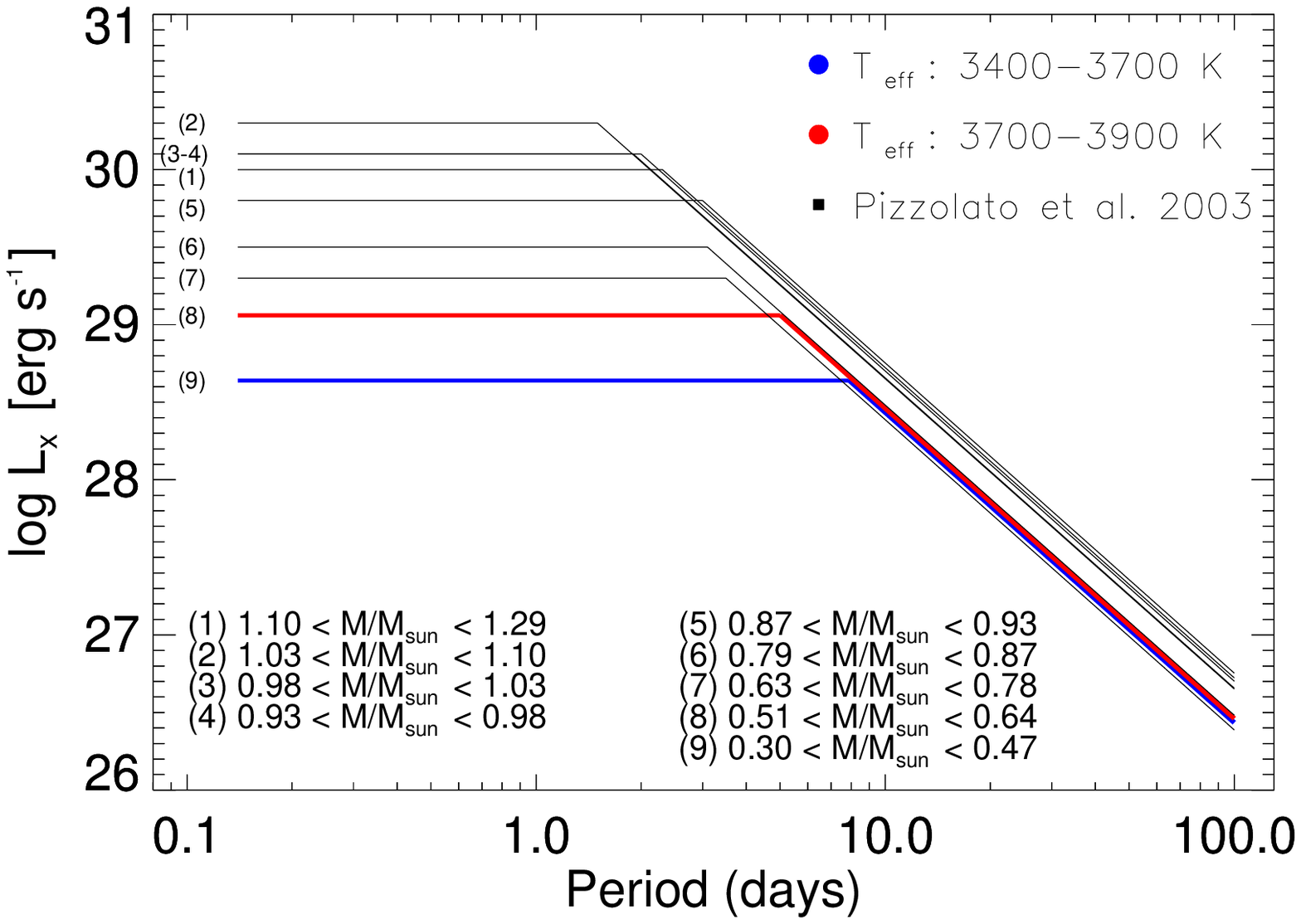}
\end{minipage}
\begin{minipage}{0.48\linewidth}
\includegraphics[angle=0,scale=0.5]{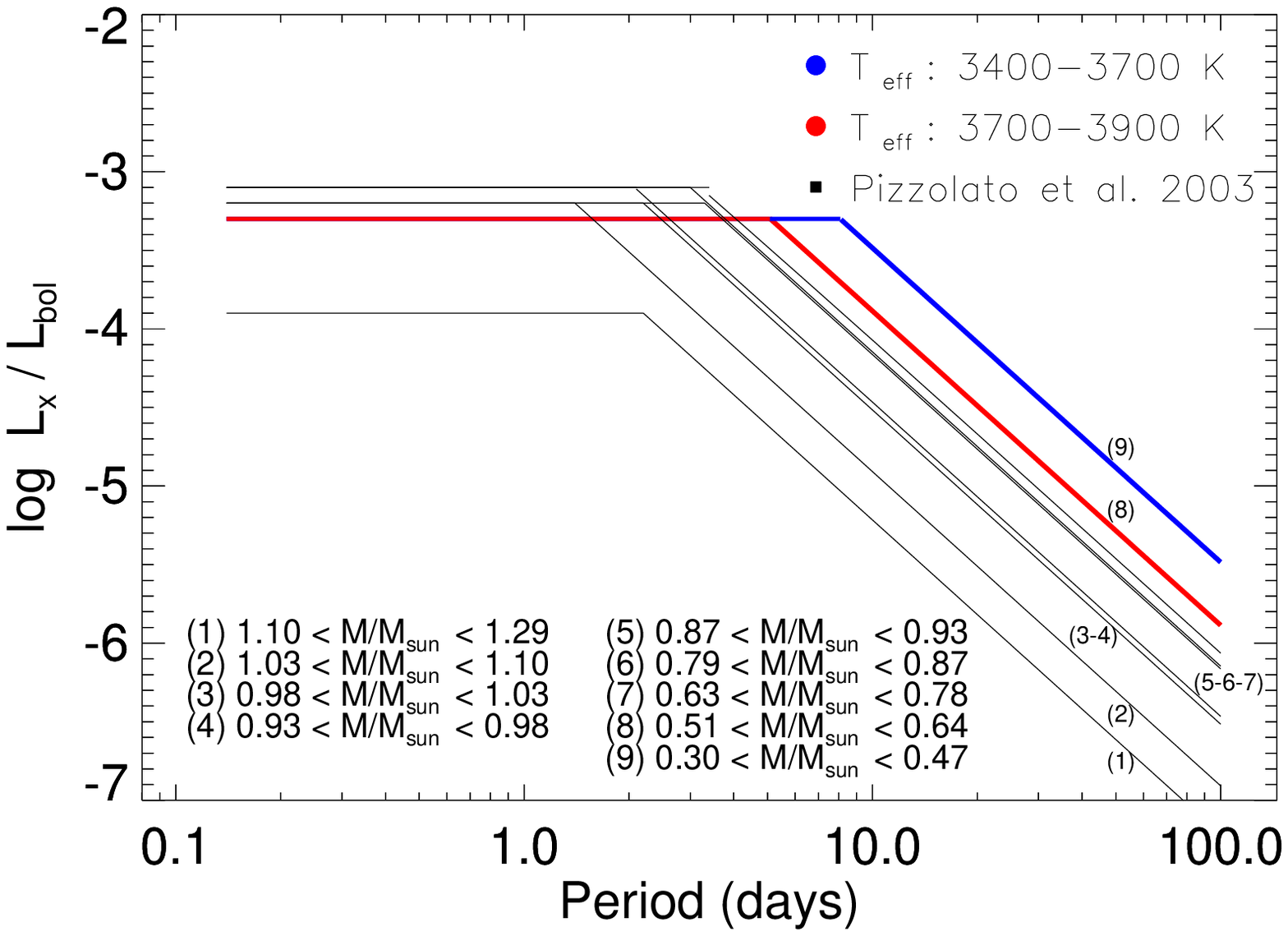}
\end{minipage}
\caption[All the best-fit relations between X-ray emission and rotational period.]{All the best-fit relations between X-ray emission and rotational period for all the mass range (F to K stars) considered in \cite{2003A&A...397..147P} are presented with black lines. The blue and red solid lines represent our best-fit for 3400-3700 and 3700-3900 K $T_{\rm eff}$ range, respectively.}
\label{fits_Lx_and_Lbol_vs_prot}
\end{figure*}

In Fig. \ref{Lx_and_Lbol_prot_all} (left panel), our data show a spread around the best fit relation by \cite{2003A&A...397..147P}. This spread is more evident in the right panel of Fig. \ref{Lx_and_Lbol_prot_all} where $\log L_x/L_{\rm bol}$ is reported. It could be due to two possible reasons: i) the mass range studied by \cite{2003A&A...397..147P} ($M$ = 0.22-0.60 $\rm M_{\odot}$) is too large and it should be divided in subgroups; ii) the derived rotational periods are too uncertain and their inclusion produces an artificially large spread. 

On the one hand we therefore excluded from our analysis the 14 derived $P_{\rm rot}$ values (i.e, those derived from calcium measurements using a Ca~{\sc ii} H \& K - $P_{\rm rot}$ relationship) maintaining only those with a direct measurement of the rotational period. We note from now that only periods measured from time series will be included in the analysis. On the other hand we divided our sample into two $T_{\rm eff}$ groups:
3400 $\leq$ $T_{\rm eff}$ $\leq$ 3700 and 3700 $\leq$ $T_{\rm eff}$ $\leq$ 3900 K (approximately mass range $M=0.3-0.47~M_{\odot}$ and $M=0.51-0.64~M_{\odot}$, respectively). This leaves us with a total of 19 M dwarfs in our sample, 11 stars in the 3400-3700 K range and 8 in the 3700-3900 K range. Indeed, after discarding stars without a direct rotation period measurement, the correlations show a remarkably smaller dispersion (Fig.  \ref{Lx_and_Lbol_vs_prot}), indicating that we have identified the origin of the spread in Fig. \ref{Lx_and_Lbol_prot_all}.

Fig. \ref{Lx_and_Lbol_vs_prot} shows the behaviour of M dwarfs in the non-saturated regime by dividing our stars into two $T_{\rm eff}$ groups plotted with different colours. We have not enough stars to determine here the saturated regime level (fast rotators) because almost all the stars in our sample are placed into the non-saturated regime. We assume the saturation level as described by \cite{2003A&A...397..147P}. In a more recent work and using a larger sample, \cite{2016csss.confE..62S} determine a higher saturation level than the one found by \cite{2003A&A...397..147P}. However, the sample in \cite{2016csss.confE..62S} is taken from the $Kepler$ Two-Wheel (K2) light curves and it is likely biased towards bright X-ray M dwarfs which can be translated into a higher value of the saturation level.

In order to obtain a new parametrization of the X-ray emission vs. rotation relationship for M dwarfs in the non-saturated regime for the two bins of $T_{\rm eff}$, the data were fitted in each temperature bin with a fixed power law exponent of -2, leaving the intercept parameter free to vary. This value follows the well studied and known relation \citep[e.g.][]{2003A&A...397..147P,2014ApJ...794..144R} for the non-saturated regime, $L_x \propto P^{-2}$, indicating that the rotation period alone determines the X-ray emission. Power-law functions were fitted to the data,

\begin{equation}
\log F_1=a_0+a_1~\log P_{rot}
\end{equation}
where $\log F_1$ corresponds with the values of $\log L_x$ and $\log L_x/L_{\rm bol}$ and $a_0$ and $a_1$ are the fit coefficients. In our case we set the value $a_1$ fixed to -2 and obtained the best two intercept parameters presented in Table \ref{tab:fit_coeff_xray_set-2}.

\begin{table}[!t]
\centering
\caption[Coefficients of the activity-rotation relationships]{Coefficients of the activity-rotation relationships with slope set to -2.}
\label{tab:fit_coeff_xray_set-2}
\begin{tabular}{c c c c c}
\hline
\hline
\noalign{\smallskip}

$T_{\rm eff}$	&	$\log F_1$ 	&	$a_0$	& $\chi^2 $\\

\noalign{\smallskip}
\hline
\noalign{\smallskip}

3400-3700 K	&	$\log L_x$ (erg $\rm s^{-1}$)				&	$30.43 \pm 0.04$		&	30.98\\

		&	$\log L_x/L_{\rm bol}$				&	$-1.48 \pm 0.06$		&	13.90\\
		
\noalign{\smallskip}		
\hline
\noalign{\smallskip}		
		
3700-3900 K	&	$\log L_x$	 (erg $\rm s^{-1}$)		&	$30.46 \pm 0.04$		&	207.09\\
		&	$\log L_x/L_{\rm bol}$			&	$-1.89 \pm 0.05$		&	96.75\\

\hline
\end{tabular}
\end{table}

The obtained best-fit relations between $\log L_x$ and $P_{\rm rot}$ (Fig. \ref{Lx_and_Lbol_vs_prot}, left panel) are very similar for the two $T_{\rm eff}$ bins. Note the complementary behaviour of the $\log L_x/L_{\rm bol}$ versus $P_{\rm rot}$ relation (Fig. \ref{Lx_and_Lbol_vs_prot}, right panel) where 
the different loci occupied by the two $T_{\rm eff}$ groups are more evident. This behaviour is coherent with the extension of \cite{2003A&A...397..147P} for more massive stars (F to late-K stars) presented in Fig. \ref{fits_Lx_and_Lbol_vs_prot}, being  our work an extension to the low mass regime that follows the observed trend for massive stars. 

Using $\log L_x/L_{\rm bol}$ as indicator, for low-mass stars that rotate very fast the relationship breaks down when X-ray luminosity reaches the approximately saturated value of $\log L_x/L_{\rm bol} \sim -3$ \citep[e.g.][]{1984A&A...133..117V,2003A&A...397..147P} independent of the stellar mass (except for the most massive bin). This saturation level is reached at rotation periods that increase towards less massive stars (increasing with decreasing the bolometric luminosity).

On the contrary, if the $\log L_x$ indicator is considered, the $\log L_x$ is a function of rotation period in the non-saturated regime being independent from stellar mass.

This behaviour, across all spectral types, is clearly seen in Fig. \ref{fits_Lx_and_Lbol_vs_prot} where we represent the collection of all best-fit relations found by \cite{2003A&A...397..147P} in all the mass ranges (F to late-K stars) including the new relations found in this work in the non-saturated regime for early-M dwarfs. In the left panel our best-fits are placed in the same range of non-saturated regime of all spectral types while in the right panel ($\log L_x/L_{\rm bol}$ vs. $P_{\rm rot}$) the new curve of the M dwarfs is on the right of the curves of more massive stars. 

We have found a continuous shift of the $\log L_x/L_{\rm bol}$ vs. $P_{\rm rot}$ power-law towards longer $P_{\rm rot}$ values extending the up to now available sample of M dwarf stars to the non-saturated X-ray emission regime with respect to previous studies \citep[e.g.][]{2003A&A...397..147P, 2016csss.confE..62S,2011ApJ...743...48W}.

Consequently we are able to determine in a more accurate way than in previous works the value of rotation period at which the saturation occurs $(P_{\rm sat})$ for M dwarf stars. Assuming a defined saturated value for all mass range in $\log L_x/L_{\rm bol}$ as -3.3 (except for the most massive bin that, as shown in the right panel of Fig. \ref{fits_Lx_and_Lbol_vs_prot}, does not reach the same saturated value of the less massive bins) and knowing the mean value of $\log L_{\rm bol}$ (for each temperature group) it is possible to estimate a better value for the rotation period $P_{\rm sat,L_x/L_{\rm bol}}$ for each temperature bin in $\log L_x/L_{\rm bol}$ representation. Then we are able to obtain the correspondence of saturated rotation period value ($P_{\rm sat,L_x}$) in $\log L_x$ representation. The rotation period values derived from our sample at which X-ray emission reaches the saturation level are presented in Table \ref{tab:satured_level}, as well as the values found in literature for early/mid M dwarfs.

\begin{table*}[!t]
\centering
\caption[X-ray saturated level for M dwarfs]{X-ray saturated level for M dwarfs.}
\label{tab:satured_level}
\begin{tabular}{c c c c c c c c}
\hline
\hline
\noalign{\smallskip}

Spectral type   & N*    &  $\log L_{x, \rm sat}$	& $\log L_{x, \rm sat}/L_{\rm bol}$ & $P_{\rm sat,L_x}$	& $P_{\rm sat,L_x/L_{\rm bol}}$ \\

	  &		  & (erg $\rm s^{-1}$)  	& 	& (days)		&	(days)\\
	
\noalign{\smallskip}
\hline
\noalign{\smallskip}

3400-3700 K ($\rm M1.0-M3.0)^{(a)}$	& 8 		&  $28.64 \pm 0.2^{(a)}$	&	$-3.3 \pm 0.2^{(c)}$	& 8.04$^{(a)}$ & 8.13$^{(a)}$\\

3700-3900 K ($\rm M0-M1.5)^{(a)}$	& 11 		&  $29.06 \pm 0.2^{(a)}$	&	$-3.3 \pm 0.3^{(c)}$	& 5.08$^{(a)}$ & 5.07$^{(a)}$\\

\noalign{\smallskip}
\hline
\noalign{\smallskip}

2900-3680 K $(\rm M2-M5.5)^{(c)}$	    	& 21 	&  $28.2 \pm 0.2$	&	$-3.3 \pm 0.2$	&	> 10.8	&	>13.1\\
$\rm K7-M2^{(b)}	$		& 5 		&  $29.2 \pm 0.4$	&	$-3.0 \pm 0.4$	& <10 & --\\
$\rm M3-M4^{(b)}	$		& 7 		&  $28.6 \pm 0.3$	&	$-3.1 \pm 0.2$	& <10	& --\\
\noalign{\smallskip}
\hline
\noalign{\smallskip}

\multicolumn{6}{l}{\footnotesize{$^{(a)}$ This work; $^{(b)}$ \cite{2016csss.confE..62S}; $^{(c)}$\cite{2003A&A...397..147P}}}
\end{tabular}
\end{table*}

\begin{figure}[!t]
\centering
\includegraphics[angle=0,scale=0.5]{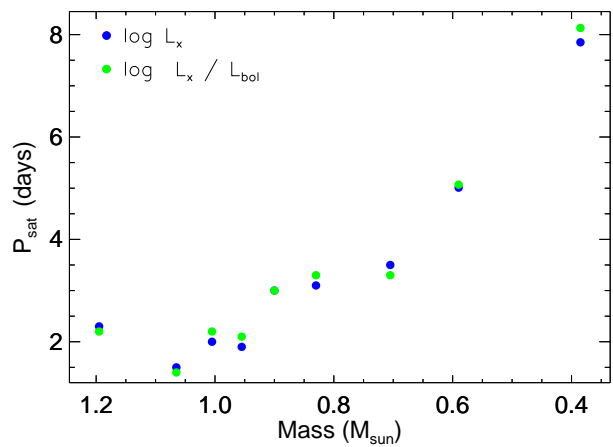}
\caption{Rotation period at which the saturation level $(P_{\rm sat})$ as a function of the stellar mass is reached. The blue and green dots correspond with the rotation period value at which the saturation occurs in $L_x$ and $L_x/L_{\rm bol}$ representation, respectively.}
\label{Prot_sat_vs_massgroup}
\end{figure}

In Fig. \ref{Prot_sat_vs_massgroup} we show the rotation period at which the saturation level is reached, $P_{\rm sat}$, both in $L_x$ and $L_x/L_{\rm bol}$ representation, as a function of the different mass ranges used previously (Fig. \ref{fits_Lx_and_Lbol_vs_prot}). The global behaviour of the $P_{\rm sat}$ value along the different masses, in both cases, is to increase towards smaller stellar masses. Therefore less massive targets will reach the saturated value at longer rotation periods. 

\section{Summary and conclusions}
\label{sec:Summary and conclusions}

The aim of this paper was to test if the known relations in previous works between activity and stellar rotation for main-sequence FGK stars also hold for our sample of early-M dwarfs. Computing the coronal activity indicator X-ray luminosity using the available data from $ROSAT$ and $XMM$-$Newton$ and using the rotation periods determined inside the HADES collaboration we were able to study the relation $L_{x} \propto P_{\rm rot}^{-2}$, that describes the non-saturated regime. We conclude that the relation is also valid for our targets extending the up to now available sample of M dwarfs to the non-saturated X-ray emission regime. Our best-fit relations follow the observed trend for more massive stars in both representations, $L_{x}$ and $L_{x}/L_{\rm bol}$ versus $P_{\rm rot}$ which let us determine in a more accurate way than in previous works the calculated value at which the rotation period achieves the saturation. This was possible by knowing the defined value of $\log L_{x}/L_{\rm bol}=-3.3$ \citep{2003A&A...397..147P} when the stars saturate, the mean value of $\log L_{\rm bol}$ for each temperature group  and considering the coefficients of the activity-rotation relationship estimated in this work.

We wanted to go one step further and study the age at which is reached the saturation level in M dwarfs. We used the angular momentum evolution models for 0.8 $M_{\odot}$ and 0.3 $M_{\odot}$ stars presented by \cite{2007IAUS..243..231B} in Fig. 4 of that work. Following the models for slow and fast rotators showed by red solid lines it is possible an estimation of the age at which the saturation level is reached. The rotational period is easily converted to angular velocity by $\Omega = 2\pi / P_{\rm rot}$ and scaled to the Sun ($\Omega_{\odot} = 2.87 \times 10^{-6}$ $\rm s^{-1}$). On the one hand, for our sample in the mass range group $\sim 0.3 M_{\odot}$ that reaches the saturation at $P_{\rm sat,L_x}\sim$ 8 days we can used the 0.3 $M_{\odot}$ models, finding that the saturation age is from 300 to 400 Myr. On the other hand, for the stars corresponding to those studied by \cite{2003A&A...397..147P}, it is better to use the models computed for 0.8 $M_{\odot}$ because of the different mass range. So, the age at which they reach the saturation level could be defined before of 300 Myr by using the 0.8 $M_{\odot}$ models. The age range depends on the exact path followed by the star during its rotational evolution.

Finally, the importance of analysing the rotation period-activity relation using our sample of M dwarfs and concluding that is also valid for our targets, lies in that our data fill the slow rotation area. This area exactly corresponds with a specific region in which planets in habitable zone around low-mass stars could be discovered.

This confirms the big challenge that the stellar activity poses on our radial velocity survey on discovering planets in early-M dwarfs and our interest on a good understanding of the stellar activity.


\begin{acknowledgements}

This work was supported by WOW from INAF through the \textit{Progetti Premiali} funding scheme of the Italian Ministry of Education, University, and Research.
This research has made use of data and/or software provided by the High Energy Astrophysics Science Archive Research Center (HEASARC), which is a service of the Astrophysics Science Division at NASA/GSFC and the High Energy Astrophysics Division of the Smithsonian Astrophysical Observatory. M.P. and I.R. acknowledge support from the Spanish Ministry of Economy and Competitiveness (MINECO) and the Fondo Europeo de Desarrollo Regional (FEDER) through grant ESP2016-80435-C2-1-R, as well as the support of the Generalitat de Catalunya/CERCA program. J.I.G.H, RR and B.T.P acknowledge financial support from the Spanish Ministry project MINECO AYA2017-86389-P. B.T.P. acknowledges Fundaci\'on La Caixa for the financial support received in the form of a Ph.D. contract. J.I.G.H. acknowledges financial support from the Spanish MINECO under the 2013 Ram\'on y Cajal program MINECO RYC-2013-14875. A.S.M acknowledges financial support from the Swiss National Science Foundation (SNSF). G.S. acknowledges financial support from ''Accordo ASI–INAF'' No. 2013-016-R.0 July 9, 2013 and July 9, 2015. A.S.M acknowledges financial support from the Swiss National Science Foundation (SNSF). MPi gratefully ackowledges the support from the European Union Seventh Framework Programme (FP7/2007-2013) under Grant Agreement No. 313014 (ETAEARTH).

\end{acknowledgements}

%
%

\bibliographystyle{aa} 
\bibliography{bibliography.bib} 




\end{document}